\documentclass[prc,twocolumn,aps,superscriptaddress,showpacs,fleqn]{revtex4}
\usepackage{amssymb}
\usepackage{amsmath,bm}
\usepackage{graphicx}
\usepackage[normalem]{ulem}
\usepackage{booktabs}
\usepackage{threeparttable}
\usepackage[dvips]{color}
\usepackage{leftidx}
\usepackage{CJK}

\renewcommand\sout{\bgroup \color{red} \ULdepth=-.5ex \ULset}

\begin{document}

\title{Positioning the neutron drip line and the r-process paths in the nuclear landscape}

\author{Rui Wang}
\affiliation{Department of Physics and Astronomy and Shanghai Key Laboratory for
Particle Physics and Cosmology, Shanghai Jiao Tong University, Shanghai 200240, China}
\author{Lie-Wen Chen\footnote{%
Corresponding author (email: lwchen$@$sjtu.edu.cn)}}
\affiliation{Department of Physics and Astronomy and Shanghai Key Laboratory for
Particle Physics and Cosmology, Shanghai Jiao Tong University, Shanghai 200240, China}
\affiliation{Center of Theoretical Nuclear Physics, National Laboratory of Heavy Ion
Accelerator, Lanzhou 730000, China}
\date{\today}

\begin{abstract}

Exploring nucleon drip lines and astrophysical rapid neutron
capture process~(r-process) paths in the nuclear landscape is extremely challenging
in nuclear physics and astrophysics. While various models predict similar proton drip line,
their predictions for neutron drip line and the r-process
paths involving heavy neutron-rich nuclei exhibit a significant variation which hampers
our accurate understanding of the r-process nucleosynthesis mechanism. Using microscopic
density functional theory with a representative set of non-relativistic
and relativistic interactions, we demonstrate for the first time that
this variation is mainly due to the uncertainty of nuclear matter symmetry energy
$E_{\rm{sym}}(\rho_{\rm{sc}})$ at the subsaturation
cross density $\rho_{\rm{sc}}=0.11/0.16\times\rho_0$~($\rho_0$ is saturation density),
which reflects the symmetry energy of heavy nuclei. Using the
recent accurate constraint on $E_{\rm{sym}}(\rho_{\rm{sc}})$ from the binding energy
difference of heavy isotope pairs, we obtain quite precise predictions for the location
of the neutron drip line, the r-process paths and the number of bound nuclei in the nuclear landscape.
Our results have important implications on extrapolating the properties of unknown neutron-rich rare
isotopes from the data on known nuclei.

\end{abstract}


\maketitle

\emph{1. Introduction.}---%
The determination of the location of neutron and proton drip lines in
the nuclear landscape is a fundamental question in nuclear physics. The drip
lines tell us what is the limit of the nuclear stability against nucleon
emission and how many bound nuclei can exist in the nuclear chart~\cite{Tho04}.
The quest for the neutron drip line (nDL) is also important for
understanding the astrophysical rapid neutron capture process (r-process)
which occurs along a path very close to the nDL in the nuclear
landscape and provides a nucleosynthesis mechanism for the origin of more
than half of the heavy nuclei in the Universe~\cite{91Cow,Lan01,Qian03,Arn07}.
While the proton drip line (pDL) has been determined up to Protactinium (proton number
$Z=91$)~\cite{NNDC}, there has little experimental information on the nDL for
$Z>8$~\cite{Bau07}.
Since the majority of rare isotopes inhabiting
along the nDL and the r-process paths are unlikely to be observed in the
terrestrial laboratory, their information has to rely on the
model extrapolation based on the known nuclei, which is so far largely uncertain
and hampers our accurate understanding of the r-process nucleosynthesis
mechanism~\cite{Gor92,Kra93,Sun08,Sch13}. To understand and reduce
the uncertainty of the model extrapolation from the known nuclei to
the unknown neutron-rich rare isotopes is thus of critical importance, and
we show here the symmetry energy plays a key role in this issue.

The nucleon drip lines are determined by nucleon separation energy of
nuclei and theoretically they can be obtained from either macroscopic
models~\cite{Mol95,Duf95,Oya10,Wan14} or microscopic density functional theory
(DFT)~\cite{Dob84,Hir97,Dob02,Gen05,Gor09,Del10,Erl12,Erl13,Afa13,Agb14}.
Although these theoretical approaches have achieved remarkable success in
describing the data on known nuclei, extrapolations to unknown nuclei appears
less certain. Different approaches or interactions, which predict similar
pDL, may give quite different predictions for the
position of the nDL especially involving heavy neutron-rich nuclei~\cite{Erl12,Erl13,Afa13,Agb14}.
Since the nuclei close to the nDL have extremely large isospin
values, the model dependence is very likely related to the poorly known nuclear matter
symmetry energy $E_{\rm{sym}}\left(\rho\right)$, which characterizes the isospin
dependent part of the equation of state (EOS) of asymmetric nuclear matter and is
a key quantity to reflect the isovector properties of nuclear effective
interactions (see, e.g., Ref.~\cite{LCK08}).
Indeed, Oyamatsu~{\it et al.}~\cite{Oya10} found a correlation between the
nDL location and the density slope $L(\rho_0)$ of the
symmetry energy at saturation density $\rho_0$. However, a recent work
by Afanasjev~{\it et al.}~\cite{Afa13} (see also Ref.~\cite{Agb14}) provided no
evidence for such a correlation, leaving a puzzle in the community.
In this work, we demonstrate that the nDL location for heavy elements is actually correlated strongly by the
magnitude of the symmetry energy at the subsaturation cross density~(scaled by $\rho_0$) $\rho_{\rm{sc}}=0.11/0.16\times\rho_0$, i.e., $E_{\rm{sym}}(\rho_{\rm{sc}})$. In particular, the recent accurate
constraint on $E_{\rm{sym}}(\rho_{\rm{sc}})$ allows us to predict quite
precisely the location of the nDL and thus the r-process paths as well as the number of bound nuclei
in the nuclear landscape.

\emph{2. The symmetry energy and drip lines.}---%
The symmetry energy plays multifaceted roles in nuclear physics and
astrophysics~\cite{LCK08,Tsa12,Lat12,Don94,Dea02} as well as new
physics beyond the standard model~\cite{Hor01b}, and it
is defined as
$E_{\rm{sym}}(\rho) =\frac{1}{2!}\frac{\partial ^{2}E(\rho ,\delta)}{\partial \delta ^{2}}|_{\delta=0}$
via an expansion of the nucleon specific energy (i.e., EOS) in an asymmetric nuclear matter, i.e.,
$E(\rho ,\delta )=E_{0}(\rho )+E_{\mathrm{sym}}(\rho )\delta ^{2}+O(\delta^{4})$
where $\rho $ is nucleon density and $\delta=(\rho _{n}-\rho _{p})/(\rho _{p}+\rho _{n})$
is the isospin asymmetry. The $E_0(\rho )$ represents the EOS of symmetric
nuclear matter and is usually expanded around $\rho_0$ as
$E_{0}(\rho ) = E_{0}(\rho _{0}) + \frac{K_0}{2!}(\frac{\rho -\rho _{0}}{3\rho _{0}})^2 + O((\frac{\rho -\rho _{0}}{3\rho _{0}})^{3})$
where the ${K_0}$ is the so-called incompressibility coefficient.
The symmetry energy $E_{\rm{sym}}(\rho)$ can be expanded around a
reference density $\rho_{\rm{r}}$ as
\begin{equation}
%\begin{split}
E_{\rm{sym}}(\rho)
%&=\frac{1}{2!}\frac{\partial ^{2}E(\rho ,\delta)}{\partial \delta ^{2}}\Big|_{\delta=0}\\
=E_{\rm{sym}}(\rho_{\rm{r}})+L(\rho_{\rm{r}})\chi_{\rm{r}} + O(\chi_{\rm{r}} ^{2}),
\label{SymmetryEnergy}
%\end{split}
\end{equation}
with $\chi_{\rm{r}} =\frac{\rho-\rho_{\rm{r}}}{3\rho_{\rm{r}}}$.
The coefficient $L(\rho_{\rm{r}})$ denotes the density slope
of the symmetry energy at $\rho_{\rm{r}}$.

In the nuclear chart, all nuclei that can exist are bounded by the neutron and
proton drip lines. Whether a nucleus can exist is determined by its single-nucleon
and two-nucleon separation energy. Since the two-nucleon drip lines usually are
more extended than the single-nucleon drip lines due to the pairing
effect, in this work we thus mainly focus on the two-neutron (-proton) separation
energy $S_{\rm{2n}}$ ($S_{\rm{2p}}$) of even-even nuclei and the corresponding
two-neutron(-proton) drip line.
The two-neutron (-proton) drip line location $N_{\rm{drip}}$ ($Z_{\rm{drip}}$)
is recognized as the neutron (proton) number of the heaviest
bound even-even nucleus within an isotope (isotone) chain which satisfy $S_{\rm{2n}} >0$ ($S_{\rm{2p}} >0$).
It should be mentioned that there could exist a secondary and even a tertiary drip line
for an isotope chain~\cite{Erl12,Erl13,Afa13,Agb14,ZhaPeiXu13} about which
we do not consider in this work.

A qualitative preview about the two-nucleon drip lines can be obtained from
the semi-empirical nuclear mass formula in which the binding energy of a nucleus with
$N$ neutrons and $Z$ protons~($A=N+Z$) is expressed as
\begin{eqnarray}
B(N,Z)&=&a_{\rm{vol}}A+a_{\rm{surf}}A^{2/3}+a_{\rm{sym}}(A)\frac{(N-Z)^2}{A} \notag \\
&+&a_{\rm{coul}}\frac{Z(Z-1)}{A^{1/3}}+E_{\rm{pair}},
\label{LDM}
\end{eqnarray}
where $a_{\rm{vol}}$, $a_{\rm{surf}}$ and $a_{\rm{coul}}$ are constants,
$E_{\rm{pair}}$ represents the pairing contribution, and $a_{\rm{sym}}(A)$ is
the symmetry energy coefficient of finite nuclei.
For a typical heavy nuclei around the nDL, such as
$^{222}$Er ($Z=68$), assuming $a_{\rm{sym}}(A+2)\approx a_{\rm{sym}}(A)$,
one can then obtain
$S_{\rm{2n}} \approx -2a_{\rm{vol}} - 0.22 a_{\rm{surf}} - 1.24 a_{\rm{sym}}(A) +2.27 a_{\rm{coul}}$.
The pairing term is eliminated for even-even nuclei. Empirically, the values of $a_{\rm{vol}}$, $a_{\rm{surf}}$ and $a_{\rm{coul}}$ are
relatively well determined, and thus the uncertainty of
$a_{\rm{sym}}(A)$ (within a few MeV) essentially dominates
the uncertainty of $S_{\rm{2n}}$ at large $N-Z$ where the nDL is concerned
and thus causes the uncertainty of the $N_{\rm{drip}}$.
Similarly, for a typical heavy nuclei around the pDL,
such as $^{222}$Cm ($Z=96$), one has
$S_{\rm{2p}} \approx -2a_{\rm{vol}} - 0.22 a_{\rm{surf}} + 0.60 a_{\rm{sym}}(A) -58.07 a_{\rm{coul}}$.
Since the pDL is close to the symmetry axis of $N=Z$, so
$a_{\rm{sym}}(A)$ would not have a significant effect on $S_{\rm{2p}}$
and thus the pDL. In addition, the Coulomb energy makes $S_{\rm{2p}}$
vary very rapidly with $Z$ and thus leads to a relatively stable $Z_{\rm{drip}}$.

The above simple argument based on the mass formula thus
indicates the $a_{\rm{sym}}(A)$ indeed plays a central role
for locating the nDL. For
heavy nuclei, considering the empirical correspondence between
$a_{\rm{sym}}(A)$ and $E_{\rm{sym}}(\rho_{\rm{c}})$~\cite{Cen09,Dan14,Che11,Fat14}
with $\rho_{\rm{c}} \approx 0.11$ fm$^{-3}$ roughly corresponding
to the nuclear average density, one then expects
$E_{\rm{sym}}(\rho_{\rm{sc}})$~(here $\rho_{\rm{c}}$ is replaced by
$\rho_{\rm{sc}}$ to consider the saturation
density variation in various models) should be strongly correlated with the
$N_{\rm{drip}}$ for heavy elements.

\emph{3. Correlation analysis.}---%
The present large-scale calculations of the nuclear binding energy are
based on the Skyrme-Hartree-Fock-Bogolyubov (SHFB) approach using the
code HFBTHO~\cite{HFBcode} and the relativistic mean field (RMF)
approach using the code DIRBHZ~\cite{Nik14}.
These codes allow for an accurate description
of deformation effects and pairing correlations in nuclei arbitrarily
close to the nucleon drip lines.
In particular, we use a large harmonic basis
corresponding to $20$ major shells and restrict ourselves to axially deformed nuclei.
In the SHFB calculations, the density-dependent $\delta $ pairing force with a
mixed-type pairing is used and the pairing strength is adjusted to fit the empirical
value of $1.245$ MeV for the neutron pairing gap of $^{120}$Sn~\cite{Dob02}.
In DIRBHZ calculations, a separable version of finite range Gogny (D1S) pairing
force is used~\cite{Nik14}.
In the following, we choose the Erbium isotope chain ($Z=68$) and
isotone chain of $N=126$ as examples to evaluate $N_{\rm{drip}}$ and
$Z_{\rm{drip}}$, respectively, through which we hope to examine the
correlation of the drip lines with different macroscopic
quantities. The choice of isotope or isotone chain is more or less
arbitrary, and the above choice is adopted to avoid the shell
effect on the drip lines, about which we will discuss later.

\begin{figure}[!htb]
\centering
\includegraphics[width=8.0cm]{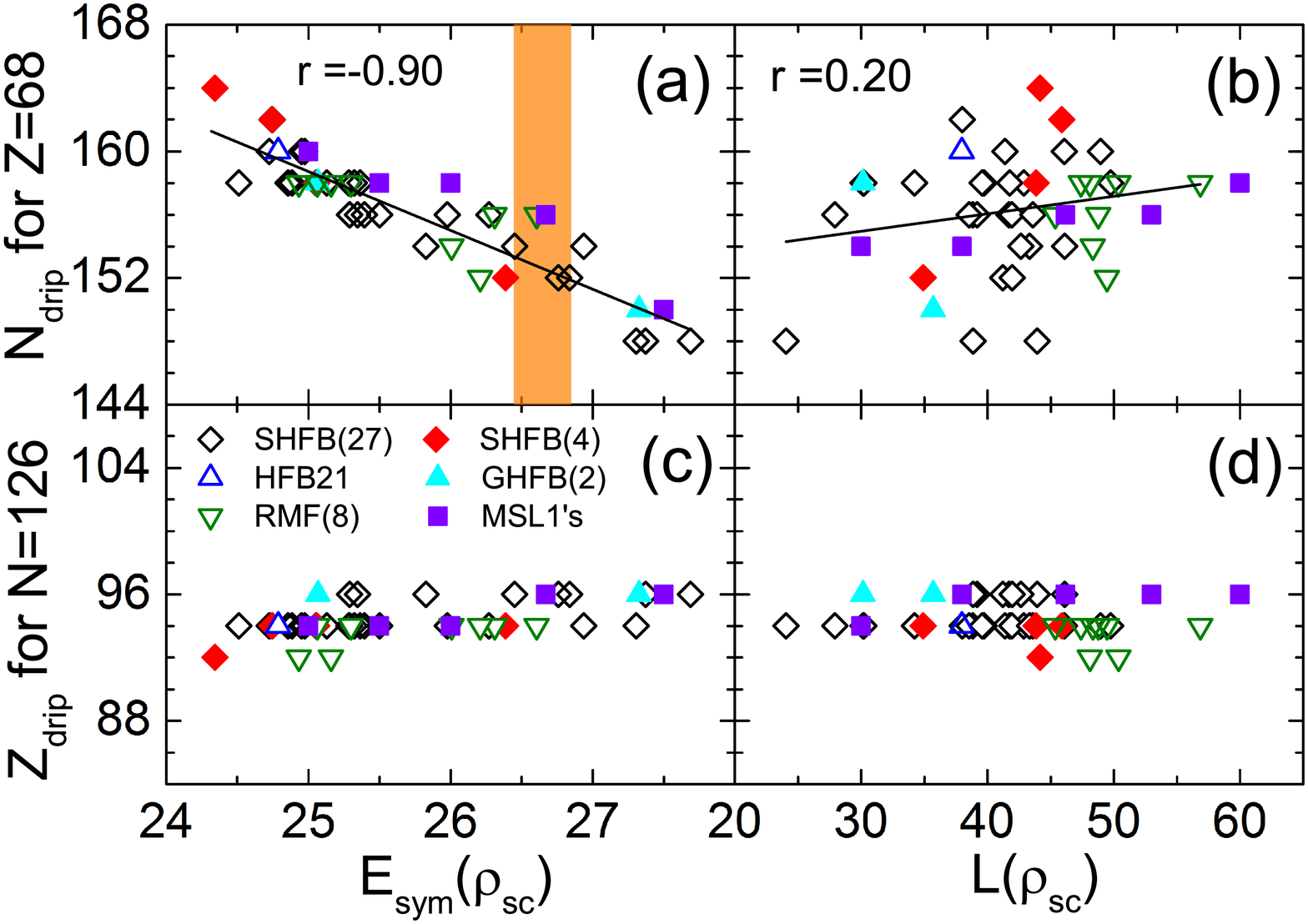}
\caption{(Color online) The calculated $N_{\rm{drip}}$ for $Z=68$ and $Z_{\rm{drip}}$
for $N=126$ versus $E_{\rm{sym}}(\rho_{\rm{sc}})$ and $L(\rho_{\rm{sc}})$.
Solid squares are the results from SHFB calculations with MSL1 by
varying individually $E_{\rm{sym}}(\rho_{\rm{sc}})$ and $L(\rho_{\rm{sc}})$.
The results from other $42$ non-relativistic and
relativistic interaction are also included. The band in (a)
indicates $E_{\rm{sym}}(\rho_{\rm{sc}}) =26.65\pm0.2$ MeV~\cite{Zha13}.
See text for details.}
\label{DLEsymRc}
\end{figure}

For the standard Skyrme interaction~(see, e.g., Ref.~\cite{Cha98}),
the nine parameters $t_{0}$-$t_{3}$, $x_{0}$-$x_{3}$ and $\sigma$
can be expressed analytically in terms of nine macroscopic quantities
$\rho_0$, $E_{0}(\rho _{0})$, ${K_0}$, $E_{\rm{sym}}(\rho_{\rm{r}})$,
$L(\rho_{\rm{r}})$, the isoscalar effective mass $m_{s,0}^{\ast}$, the
isovector effective mass $m_{v,0}^{\ast}$, the gradient coefficient
$G_{S}$ and the symmetry-gradient coefficient
$G_{V}$~\cite{Che10}. In such a way, one can easily
examine the correlations of nuclear structure properties with these
macroscopic quantities by varying them individually within their empirical
constrains~\cite{Che10}.
By varying $E_{\rm{sym}}(\rho_{\rm{sc}})$ while keeping other
quantities, i.e., $\rho _0$, $E_0(\rho _0)$, $K_0$,
$L(\rho_{\rm{sc}})$, $m_{s,0}^{\ast }$, $m_{v,0}^{\ast }$, $G_{S}$,
$G_{V}$ and the spin-orbit coupling $W_0$ at their default values in
MSL1~\cite{Zha13}, we show in Fig.~\ref{DLEsymRc} (a) by solid squares
the $N_{\rm{drip}}$ for $Z=68$ as a function of $E_{\rm{sym}}(\rho_{\rm{sc}})$.
As expected, it is seen that the $N_{\rm{drip}}$ exhibits a strong dependence
on $E_{\rm{sym}}(\rho_{\rm{sc}})$, and it rapidly deceases with the increment of $E_{\rm{sym}}(\rho_{\rm{sc}})$.
Similar analyses indicate that the
$N_{\rm{drip}}$ is insensitive to $L(\rho_{\rm{sc}})$ and the $Z_{\rm{drip}}$
for $N=126$ displays very weak dependence on both $E_{\rm{sym}}(\rho_{\rm{sc}})$
and $L(\rho_{\rm{sc}})$ as shown by the solid squares in
Fig.~\ref{DLEsymRc} (b), (c) and (d), respectively.
Applying the similar analysis to
other macroscopic quantities, we find except that $E_0$ and $K_0$
show some effects on the $N_{\rm{drip}}$, both the $N_{\rm{drip}}$
and $Z_{\rm{drip}}$ display essentially no dependence on
all the other macroscopic quantities~\cite{WangR15}. Since $E_0$ and $K_0$
are nowadays relatively well determined in the DFT, the
$E_{\rm{sym}}(\rho_{\rm{sc}})$ thus indeed plays a decisive
role in locating the nDL.

In order to confirm the strong correlation between $E_{\rm{sym}}(\rho_{\rm{sc}})$
and $N_{\rm{drip}}$ observed from the above simple correlation analyses,
we also include in Fig.~\ref{DLEsymRc} the corresponding results with $42$
other well-calibrated non-relativistic and relativistic interactions, namely,
SHFB with $31$ Skyrme interactions
including $27$~\cite{ChenR12} (i.e., BSk14, SKM, RATP, SKT3, BSk11, BSk7, BSk10, SKT8, BSk5, SKT1,
BSk4, BSk15, SKT6, MSK1, MSK2, BSk1, SKXce, SLy8, SLy4, SLy5, KDE, SLy9, Skz0,
Z$^*_\sigma$, KDE0, Skz1 and a new Skyrme interaction MSL1$^*$~\cite{WangR15})
obtained in this work and $4$~(SV-min, UNEDF0, SKM*,
SkP) from Ref.~\cite{Erl12}, HFB with HFB21~\cite{Erl12,Gor10}, Gogny-HFB~(GHFB)
with D1S and D1M~\cite{Gor09,Del10}, and RMF model with $8$ interactions including
$5$~\cite{ChenLW07,Typ10} (DD, DD-2, DD-ME1, TW99 and DD-F) obtained in this work and $3$
(DD-PC1, DD-ME$\delta $, DD-ME2) from Ref.~\cite{Agb14}.
We select these interactions
in order to have a large spread of the $E_{\rm{sym}}(\rho_{\rm{sc}})$ values
within the empirical range of $24 - 28$ MeV~\cite{Dan14}. It is seen that the results
from these interactions indeed follow the systematics from the simple correlation analysis
above. The Pearson coefficient $r$ for the
$N_{\rm{drip}}$-$E_{\rm{sym}}(\rho_{\rm{sc}})$ correlation from the $43$
interactions~(including MSL1) is $-0.90$, and this is a pretty strong (anti-)correlation
considering the fact that the $N_{\rm{drip}}$ is varied in unit of $2$.
In addition, one can also see from Fig.~\ref{DLEsymRc} that the $N_{\rm{drip}}$
from the $43$ interactions exhibits a very weak correlation with the
$L(\rho_{\rm{sc}})$~($r=0.20$), and both $E_{\rm{sym}}(\rho_{\rm{sc}})$ and
$L(\rho_{\rm{sc}})$ essentially have no effects on the $Z_{\rm{drip}}$.

\begin{figure}[htb]
\centering
\includegraphics[width=8.4cm]{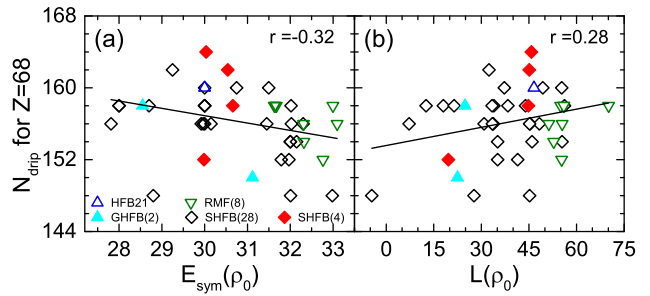}
\caption{(Color online) The calculated $N_{\rm{drip}}$ for $Z=68$ versus
$E_{\rm{sym}}(\rho_0)$ and $L(\rho_0)$ from $43$ non-relativistic and
relativistic interactions. See text for details.}
\label{DLEsymR0}
\end{figure}

It is interesting to examine the correlations of the $N_{\rm{drip}}$ for
$Z=68$ with $E_{\rm{sym}}(\rho_0)$ and $L(\rho_0)$, and the results are plotted
in Fig.~\ref{DLEsymR0} using the $43$ interactions. It is seen that
both the correlations are quite weak, i.e., the $r$ value is $-0.32$ for
$N_{\rm{drip}}$-$E_{\rm{sym}}(\rho_0)$ and $0.28$ for $N_{\rm{drip}}$-$L(\rho_0)$,
consistent with the conclusion in Refs.~\cite{Afa13,Agb14}.
This feature is due to the fact that the
$N_{\rm{drip}}$ depends on both $E_{\rm{sym}}(\rho_0)$ and $L(\rho_0)$ due to
their correlation with $E_{\rm{sym}}(\rho_{\rm{sc}})$~\cite{Che11}.
Furthermore, if $E_{\rm{sym}}(\rho_0)$
($L(\rho_0)$) is fixed, increasing $L(\rho_0)$ (decreasing $E_{\rm{sym}}(\rho_0)$)
will lead to a decrease of $E_{\rm{sym}}(\rho_{\rm{sc}})$~\cite{Che11} and
thus an increase of $N_{\rm{drip}}$,
consistent with the results in Ref.~\cite{Oya10}.

\begin{figure*}[!hbt]
\centering
\includegraphics[width=14.5cm]{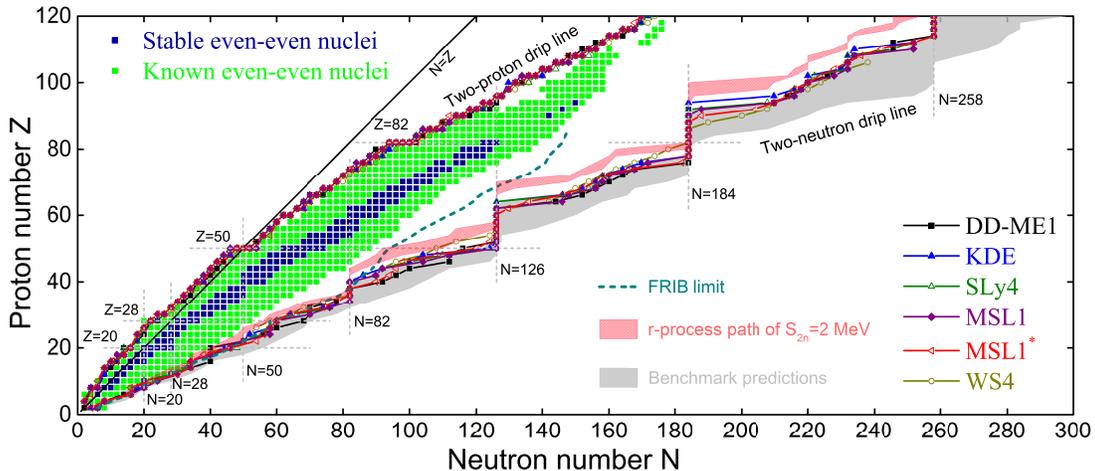}
\caption{(Color online) The landscape of bound even-even nuclei as obtained from DFT calculations with four Skyrme interactions (KDE, SLy4, MSL1, MSL1$^*$) and one relativistic interaction (DD-ME1).
The prediction from Weizsacker-Skyrme mass formula with WS4~\cite{Wan14} is also included for comparison.
The gray band denotes the uncertainty of two-neutron drip line from the benchmark calculations in Refs.~\cite{Erl12,Erl13,Afa13,Agb14}.
The red band shows the r-process path of $S_{\rm{2n}}=2~\rm{MeV}$ for KDE, SLy4, MSL1, MSL1$^*$ and DD-ME1.
The experimentally known $800$ bound even-even nuclei (up to 2014), including $169$ stable (navy squares) and $631$ radioactive (green squares), are extracted from Ref.~\cite{Tho13-15} and references therein.}
\label{NuclChart}
\end{figure*}

\emph{4. Neutron drip line and r-process path.}---%
The above analyses suggest a precise value of $E_{\rm{sym}}(\rho_{\rm{sc}})$
will put stringent constraint on the location of the nDL and the r-process paths.
Using the accurate constraint of $E_{\rm{sym}}(\rho_{\rm{sc}}) =26.65\pm0.2$ MeV
extracted recently by analyzing the binding energy difference of heavy isotope
pairs~\cite{Zha13}, one can thus obtain the drip lines by using five of the
previous $43$ interactions, i.e., KDE, SLy4, MSL1, MSL1$^*$ and DD-ME1, which
are consistent with $E_{\rm{sym}}(\rho_{\rm{sc}}) =26.65\pm0.2$ MeV
(see the band in Fig.~\ref{DLEsymRc} (a)), and the results are shown in
Fig.~\ref{NuclChart}.
Also included in Fig.~\ref{NuclChart}
are the experimentally known even-even nuclei.
The recently measured light neutron-unbound nuclei $^{16}$Be~\cite{Spy12} and
$^{26}$O~\cite{Lun12} are not included.
It is very interesting to see that these interactions indeed give quite similar
predictions for both the neutron and proton drip lines.
In light mass region where the DFT calculation is
expected to be less reliable, while KDE, SLy4, MSL1 and DD-ME1 predict
$^{26}$O or $^{28}$O to be the two-neutron drip line of Oxygen,
MSL1$^*$ predicts $^{24}$O which agrees with the current experimental
suggestion~\cite{Lun12}. In addition, we note that a small variation of other parameters can easily vary
$N_{\rm{drip}}$ by $2$ for Oxygen~\cite{WangR15}.
It should be mentioned that the lighter two-neutron drip line nucleus $^{24}$O
can also be predicted by considering the repulsive three-body force~\cite{Ots10}.

Also included in Fig.~\ref{NuclChart} are the drip lines from the Weizsacker-Skyrme
mass formula with the most recent parameter set WS4~\cite{Wan14} which predicts a
quite small rms deviation of $298$ keV with respect to essentially all the available
mass data, and they are seen to agree well with the five microscopic calculations.
We would like to point out that all the five microscopic predictions are
consistent with the benchmark calculations~\cite{Erl12,Erl13,Afa13,Agb14}
(the gray band in Fig.~\ref{NuclChart}) but with a much smaller uncertainty,
indicating the importance of an accurate value of $E_{\rm{sym}}(\rho_{\rm{sc}})$.

The precise position of the two-nucleon drip lines allows us to estimate the
number of bound even-even nuclei with $2\leqslant Z\leqslant120$, and the result
is $1887$ for KDE, $1928$ for SLy4, $1975$ for MSL1,
$1953$ for MSL1$^*$ and $1963$ for DD-ME1,
indicating a quite precise value of $1941 \pm 31$~(only 800 have been discovered experimentally~\cite{Tho13-15}).
The uncertainty mainly comes from the shell effect which will be discussed later.
Furthermore, the single-nucleon drip lines can be estimated from the condition
$-E^{\rm{F}}_{\rm{n,p}}(N,Z)=\Delta_{\rm{n,p}}(N,Z)$ while
the two-n(p)DL of odd-Z(N) nuclei can be estimated
from the condition $2E^{\rm{F}}_{\rm{n(p)}}(N,Z)=0$,
where the Fermi energy $E^{\rm{F}}$ and pairing gap $\Delta$ of odd-odd or
odd-A nuclei can be approximated by the average of the corresponding calculated
results of their even-even neighbors~\cite{Erl12}. Accordingly, we estimate the
total number of bound nuclei to be $6794$, $6895$, $7115$ and $6659$
for KDE, SLy4, MSL1 and MSL1$^*$, respectively,
leading to a precise estimate of $6866 \pm 166$~(only $3191$ have been
discovered experimentally~\cite{Tho13-15}).
Although the above candidate interactions are not a large sample,
the small variation of their predictions represents a useful estimate of the
uncertainty from sources other than $E_{\rm{sym}}(\rho_{\rm{sc}})$.

The astrophysical r-process is expected to occur along a path of constant neutron
separation energies~\cite{91Cow,Lan01,Qian03,Arn07,Gor92,Kra93,Sun08,Sch13}.
For the five candidate interactions, we also include in Fig.~\ref{NuclChart}
the r-process path of $S_{\rm{2n}}=2~\rm{MeV}$~(the results with other $S_{\rm{2n}}$
values can be provided by the authors on request). One can see the five
interactions give fairly consistent r-process paths.
Further shown in Fig.~\ref{NuclChart} is the limit of neutron-rich isotopes that
might be measured at FRIB~\cite{Afa15,FRIB2011}, and our results suggest that the future
FRIB experiment measurement may cover the nDL for $Z \lesssim 30$ and $Z \approx 40$
as well as the r-process path for $Z \lesssim 50$ and $Z \approx 70$.

As shown in Fig.~\ref{NuclChart}, the nDL exhibits a clear shell
structure, i.e., around neutron magic numbers $N= 82, 126, 184$ and $258$, the
position of the nDL is robust.
However, the $Z$ value ($Z_{\rm sh}$) at which the nDL moves away from the neutron
magic number is sensitive to the interactions, and this is the main reason
for the small variation in the predicted number of bound nuclei from
the candidate interactions as mentioned earlier.
Using a similar analysis as before, we find that the $Z_{\rm sh}$ value is
sensitive to all macroscopic quantities, i.e., $\rho _{0}$, $E_{0}(\rho _{0})$,
$K_{0}$, $m_{s,0}^{\ast }$, $m_{v,0}^{\ast }$, $G_{S}$, $G_{V}$, $W_{0}$,
$E_{\rm{sym}}(\rho_{\rm{sc}})$ and $L(\rho_{\rm{sc}})$~\cite{WangR15}, 
indicating the complexity of an accurate prediction of the $Z_{\rm sh}$ value.

\emph{5. Conclusion.}---%
In summary, using microscopic density functional theory with a number of
representative non-relativistic and relativistic interactions, we have found
a strong correlation between the neutron drip line location and the magnitude
of the symmetry energy $E_{\rm{sym}}(\rho_{\rm{sc}})$ at the subsaturation cross
density~(scaled by $\rho_0$) $\rho_{\rm{sc}}=0.11/0.16\times\rho_0$.
This finding together with the recent
accurate constraint on $E_{\rm{sym}}(\rho_{\rm{sc}})$ from the binding energy
difference of heavy isotope pairs allows us to obtain quite precise
predictions for the location of the neutron drip line, the r-process
paths and the number of bound nuclei in the nuclear landscape.
Our work sheds light on extrapolating the properties 
of unknown neutron-rich rare isotopes from the data on known nuclei.
The present results should
be less model dependent since they are based on a large
set of both non-relativistic and relativistic models. In addition,
although we have only used the lowest-order (quadratic) symmetry energy term $E_{\rm{sym}}(\rho)$ to characterize
the isospin-dependent part of the EOS for asymmetric nuclear matter,
all the higher-order symmetry energy terms have been considered self-consistently
in the mean-field calculations.

\emph{Acknowledgments.}---%
We thank S. Goriely for providing us the data of Gogny-HFB calculations,
N. Wang for WS4 data, and Z. Zhang for the help with constructing the MSL1$^*$ interaction.
We would also like to thank C. M. Ko, B. A. Li, and
W. Nazarewicz for very helpful discussions and comments.
This work was supported in part by the National Basic Research Program of
China (973 Program) under Contracts No. 2013CB834405 and No. 2015CB856904,
the NNSF of China under Grant Nos. 11275125 and 11135011, the ``Shu Guang"
project supported by Shanghai Municipal Education Commission and Shanghai
Education Development Foundation, the Program for Professor of Special
Appointment (Eastern Scholar) at Shanghai Institutions of Higher Learning,
and the Science and Technology Commission of Shanghai Municipality (11DZ2260700).


\begin{thebibliography}{99}

\bibitem{Tho04} M. Thoennessen, Rep. Prog. Phys. \textbf{67}, 1187 (2004).

\bibitem{91Cow} J.J. Cowan, F.-K. Thielemann, and J.W. Truran, Phys. Rep. \textbf{208}, 267 (1991).

\bibitem{Lan01} K. Langanke and M. Wiescher, Rep. Prog. Phys. \textbf{64}, 1657 (2001).

\bibitem{Qian03} Y.-Z. Qian, Prog. Part. Nucl. Phys. \textbf{50}, 153 (2003).

\bibitem{Arn07} M. Arnould, S. Goriely, and K. Takahashi, Phys. Rep. \textbf{450}, 97 (2007).

\bibitem{NNDC} National Nuclear Data Centre. Evaluated Nuclear Structuree Data File. http://www.nndc.bnl.gov/ensdf/.

\bibitem{Bau07} T. Baumann \textsl{et al.}, Nature \textbf{449}, 1022 (2007).

\bibitem{Gor92} S. Goriely and M. Arnould, Astron. Astrophys. \textbf{262}, 73 (1992); S. Wanajo, S. Goriely, M. Samyn, and N. Itoh, Astrophys. J. \textbf{606}, 1057 (2004).

\bibitem{Kra93} K.-L. Kratz, J.-P. Bitouzet, F.-K. Thielemann, P. Moller, and B. Pfeiffer, Astrophys. J. \textbf{403}, 216 (1993); K.-L. Kratz, K. Farouqi, and P. Moller, Astrophys. J. \textbf{792}, 6 (2014).

\bibitem{Sun08} B. Sun \textsl{et al.}, Phys. Rev. C \textbf{78}, 025806 (2008).

\bibitem{Sch13} J. Van Schelt \textsl{et al.}, Phys. Rev. Lett. \textbf{111}, 061102 (2013).

\bibitem{Mol95} P. M$\rm{\ddot{o}}$ller, J.R. Nix, W.D. Myers, and W.J. Swiatecki, At. Data Nucl. Data Tables \textbf{59}, 185 (1995).

\bibitem{Duf95} J. Duflo and A.P. Zuker, Phys. Rev. C \textbf{52}, R23 (1995).

\bibitem{Oya10} K. Oyamatsu, K. Iida, and H. Koura, Phys. Rev. C \textbf{82}, 027301 (2010).

\bibitem{Wan14} N. Wang, M. Liu, X.Z. Wu, and J. Meng, Phys. Lett. \textbf{B734}, 215 (2014).

\bibitem{Dob84} J. Dobaczewski, H. Flocard, and J. Treiner, Nucl. Phys. \textbf{A422}, 103 (1984); M.V. Stoitsov, J. Dobaczewski, W. Nazarewicz, S. Pittel, and D.J. Dean, Phys. Rev. C \textbf{68}, 054312 (2003); M.V. Stoitsov, W. Nazarewicz, and N. Schunck, Int. J. Mod. Phys. E \textbf{18}, 816 (2009).

\bibitem{Hir97} D. Hirata \textsl{et al.}, Nucl. Phys. \textbf{A616}, 438c (1997).
%D. Hirata, K. Sumiyoshi, I. Tanihata, Y. Sugahara, T. Tachibana, and H. Toki,

\bibitem{Dob02} J. Dobaczewski, W. Nazarewicz, and M.V. Stoitsov, Eur. Phys. J. A \textbf{15}, 21 (2002).

\bibitem{Gen05} L. Geng, H. Toki, and J. Meng, Prog. Theor. Phys. \textbf{113}, 785 (2005).

\bibitem{Gor09} S. Goriely, S. Hilaire, M. Girod, and S. P\'{e}ru, Phys. Rev. Lett. \textbf{102}, 242501 (2009).

\bibitem{Del10} J.-P. Delaroche \textsl{et al.}, Phys. Rev. C \textbf{81}, 014303 (2010).

\bibitem{Erl12} J. Erler \textsl{et al.}, Nature \textbf{486}, 509 (2012).
%N. Birge, M. Kortelainen, W. Nazarewicz, E. Olsen, A.M. Perhac, M. Stoitsov

\bibitem{Erl13} J. Erler, C.J. Horowitz, W. Nazarewicz, M. Rafalski,
and P.-G. Reinhard, Phys. Rev. C \textbf{87}, 044320 (2013).

\bibitem{Afa13} A.V. Afanasjev, S.E. Agbemava, D. Ray, and P. Ring, Phys. Lett. \textbf{B726}, 680 (2013).

\bibitem{Agb14} S.E. Agbemava, A.V. Afanasjev, D. Ray, and P. Ring, Phys. Rev. C \textbf{89}, 054320 (2014).

\bibitem{LCK08} B.A. Li, L.W. Chen, and C.M. Ko, Phys. Rep. \textbf{464}, 113 (2008).

\bibitem{Tsa12} B.M. Tsang \textit{et al}., Phys. Rev. C \textbf{86}, 015803 (2012).

\bibitem{Lat12} J.M. Lattimer, Ann. Rev. Nucl. Part. Sci. \textbf{62}, 485 (2012).

\bibitem{Don94} P. Donati, P.M. Pizzochero, P.F. Bortignon, and R.A. Broglia, Phys. Rev. Lett. \textbf{72}, 2835 (1994).

\bibitem{Dea02} D.J. Dean, K. Langanke, and J.M. Sampaio, Phys. Rev. C \textbf{66}, 045802 (2002).

\bibitem{Hor01b} C.J. Horowitz, S.J. Pollock, P.A. Souder, and R. Michaels, Phys. Rev. C \textbf{63}, 025501 (2001);
T. Sil, M. Centelles, X. Vi\~{n}as, and J. Piekarewicz, Phys. Rev. C \textbf{71}, 045502 (2005);
P.G. Krastev and B.A. Li, Phys. Rev. C \textbf{76}, 055804 (2007); D.H. Wen, B.A. Li, and L.W. Chen, Phys. Rev. Lett. \textbf{103}, 211102 (2009);
H. Zhang, Z. Zhang, and L.W. Chen, JCAP \textbf{08}, 011 (2014).

\bibitem{ZhaPeiXu13} Y.N. Zhang, J.C. Pei, and F.R. Xu, Phys. Rev. C \textbf{88}, 054305 (2013).

\bibitem{Cen09} M. Centelles, X. Roca-Maza, X. Vi\~{n}as, and M. Warda, Phys. Rev. Lett. \textbf{102}, 122502 (2009).

\bibitem{Che11} L.W. Chen, Phys. Rev. C \textbf{83}, 044308 (2011).

\bibitem{Dan14} P. Danielewicz and J. Lee, Nucl. Phys. \textbf{A922}, 1 (2014).

\bibitem{Fat14} F.J. Fattoyev, W.G. Newton, and B.A. Li, Phys. Rev. C \textbf{90}, 022801(R) (2014).

\bibitem{HFBcode} M.V. Stoitsov, J. Dobaczewski, W. Nazarewicz, and P. Ring, Comp. Phys. Comm. \textbf{167}, 43 (2005).

\bibitem{Nik14} T. Nik$\check{\rm{s}}$i$\acute{\rm{c}}$, N. Paar, D. Vretener, and P. Ring, Comp. Phys. Comm. \textbf{185}, 1808 (2014).

\bibitem{Cha98} E. Chahanat, P. Bonche, P. Haensel, J. Meyer, and R. Schaeffer, Nucl. Phys. \textbf{A627}, 710 (1997);
Nucl. Phys. \textbf{A635}, 231 (1998); erratum \textbf{643}, 441 (1998).

\bibitem{Che10} L.W. Chen, C.M. Ko, B.A. Li, and J. Xu, Phys. Rev. C \textbf{82}, 024321 (2010); L.W. Chen and J.Z. Gu, J. Phys. G \textbf{39}, 035104 (2012).

\bibitem{Zha13} Z. Zhang and L.W. Chen, Phys. Lett. \textbf{B726}, 234 (2013).

\bibitem{WangR15} R. Wang, Z. Zhang, and L.W. Chen, in preparartion, 2015.

\bibitem{ChenR12} R. Chen \textsl{et al.}, Phys. Rev. C \textbf{85}, 024305 (2012).
%B.J. Cai, L.W. Chen, B.A. Li, X.H. Li, and C. Xu,

\bibitem{Gor10} S. Goriely, N. Chamel, and J.M. Pearson, Phys. Rev. C \textbf{82}, 035804 (2010).

\bibitem{ChenLW07} L.W. Chen, C.M. Ko, and B.A. Li, Phys. Rev. C \textbf{76}, 054316 (2007).

\bibitem{Typ10} S. Typel, G. Ropke, T. Klahn, D. Blaschke, and H.H. Wolter, Phys. Rev. C \textbf{81}, 015803 (2010).

\bibitem{Tho13-15} M. Thoennessen, Rep. Prog. Phys. \textbf{76}, 056301 (2013); Int. J. Mod. Phys. E \textbf{23}, 1430002 (2014);
Int. J. Mod. Phys. E  \textbf{24}, 1530002 (2015).

\bibitem{Spy12} A. Spyrou \textsl{et al.}, Phys. Rev. Lett. \textbf{108}, 102501 (2012).

\bibitem{Lun12} E. Lunderberg \textsl{et al.}, Phys. Rev. Lett. \textbf{108}, 142503 (2012).

\bibitem{Ots10} T. Otsuka \textsl{et al.}, Phys. Rev. Lett. \textbf{105}, 032501 (2010).

\bibitem{Afa15} A.V. Afanasjev \textsl{et al.}, Phys. Rev. C \textbf{91}, 014324 (2015).

\bibitem{FRIB2011} FRIB Estimated Rates, Version 1.06, http://groups.nscl.msu.edu/frib/rates/fribrates.html.

\end{thebibliography}
\end{document}